\documentclass[pre,twocolumn,showkeys,amsmath]{revtex4}

\usepackage{graphicx}
\usepackage{bm}

\newcommand{\vecG}{\mathbf{\Gamma}}
\newcommand{\vecx}{\mathbf{x}}
\newcommand{\vecp}{\mathbf{p}}
\newcommand{\vecF}{\mathbf{F}}
\newcommand{\vecX}{\mathbf{X}}
\newcommand{\veca}{\mathbf{a}}
\newcommand{\lang}{\left\langle}
\newcommand{\rang}{\right\rangle}

\begin{document}

\title{Entropy and density of states from isoenergetic nonequilibrium processes}

\author{Artur B. Adib}
\email{artur@brown.edu}
\affiliation{
Theoretical Division, T-13, MS B213, Los Alamos National Laboratory, Los Alamos, New Mexico 87545, USA
}
\affiliation{
Department of Physics, Box 1843, Brown University, Providence, Rhode Island 02912, USA
}

\date{\today}

\begin{abstract}
Two identities in statistical mechanics involving entropy differences (or ratios of density of states) at constant energy are derived. The first provides a nontrivial extension of the Jarzynski equality to the microcanonical ensemble [C. Jarzynski, Phys. Rev. Lett. {\bf 78}, 2690 (1997)], which can be seen as a ``fast-switching'' version of the adiabatic switching method for computing entropies [M. Watanabe, W. P. Reinhardt, Phys. Rev. Lett. {\bf 65}, 3301 (1990)]. The second is a thermodynamic integration formula analogous to a well-known expression for free energies, and follows after taking the quasistatic limit of the first. Both identities can be conveniently used in conjunction with a scaling relation (herein derived) that allows one to extrapolate measurements taken at a single energy to a wide range of energy values. Practical aspects of these identities in the context of numerical simulations are discussed.
\end{abstract}


\maketitle


\section{Introduction}

In the ubiquitous studies of phase stability, phase transitions and reaction directions, most properties of interest are embodied in the free energies and entropies corresponding to the desired phases or reaction states. Not surprisingly, a great amount of intellectual effort has been directed towards finding ever more efficient computational means of estimating these quantities \cite{frenkel02}. Within the wider domain of equilibrium thermostatistics, a similar role is played by the so-called density of states, the knowledge of which generates not only free energies and entropies, but also heat capacities and the very partition function. Here, too, one finds a great variety of computational techniques specifically designed for its estimation (see, e.g., \cite{yan03} and references therein).

Traditionally, the aforementioned computational methods are inherently {\em nondynamical}, i.e. the relevant data is obtained from either a single equilibrium macrostate, or a series of such states. An exception is provided by the adiabatic switching method of Watanabe and Reinhardt \cite{watanabe90}, which allows one to recover entropy differences from a single dynamical trajectory connecting the states of interest. In principle, this method yields an exact estimate of the entropies provided the ``switching process'' is infinitely slow; in practice, one has to cope with the systematic errors that arise due to the unavoidable finite switching time \cite{frenkel02} (see also \cite{adib02b} for further aspects of this method). In the context of free energy differences, this situation has substantially changed after an identity connecting free energy differences at constant temperature and {\em nonequilibrium} processes was derived by Jarzynski \cite{jarzynski97-prl, jarzynski97-pre}, viz.
\begin{equation} \label{jarzynski}
  e^{-\Delta F(T)/T} = \lang e^{-W/T} \rang,
\end{equation}
where $\Delta F(T) \equiv F_B(T) - F_A(T)$ is the (Helmholtz) free energy difference between the states of interest $A$ and $B$, the angle brackets denote an average over all realizations of a predefined switching process connecting the states $A$ and $B$, and $W$ is the net amount of work performed on the system during each realization (throughout this paper, the units will be chosen so that the Boltzmann constant $k_B$ equals unity). Unlike the method of Ref.~\cite{watanabe90}, the above equality does not depend upon the rate at which the dynamical switching process is performed, and hence is free from the systematic errors associated with finite switching times. Another welcome computational aspect of this identity is that it is trivially parallelizable -- each realization of the process can be performed separately. Because of these features, the Jarzynski equality (JE) is becoming increasingly popular in the simulation community, and various reviews are already available (see e.g. \cite{frenkel02,jarzynski02-lecnotes,ritort03,park04}).

Although its generalization to free energies other than the Helmholtz one is straightforward \cite{crooks98,park04}, the analog of the JE for {\em entropies} cannot be obtained by such a direct route. In this case, one is intuitively tempted to adapt the derivation of Ref.~\cite{jarzynski97-prl} by simply replacing the Boltzmann factor with the microcanonical distribution, but this introduces a fundamental difficulty related to the finite (or infinitesimal) support of the latter\footnote{The reader familiar with the original derivation of the Jarzynski equality in Ref.~\cite{jarzynski97-prl} is invited to try this replacement by him/herself.}. 

The present work aims at providing a solution to the above difficulty by considering an initially microcanonical system that evolves in time under a {\em non-Hamiltonian, isoenergetic} dynamics\footnote{This possibility was conjectured by Chris Jarzynski (personal communication) during a presentation of Eq.~(\ref{dS-int}) by the author.}. As we shall see by explicit construction of such a dynamics, the problem associated with the microcanonical distribution is overcome, and one is able to derive a nonequilibrium identity for {\em entropy differences at constant energy}, much like Eq.~(\ref{jarzynski}) for free energy differences at constant temperature [see, in particular, Eq.~(\ref{identity-particular})]. In addition, its quasistatic limit reduces to an identity that generalizes a well-known thermodynamic integration formula for free energies [Eq.~(\ref{dS-int})]. Though admittedly difficult to realize in real experiments, the isoenergetic processes that underly these results are easily implemented numerically, as discussed shortly.

The remainder of this paper is organized as follows: In Sec.~\ref{derivation}, the central nonequilibrium identity is introduced and used to derive some interesting limiting cases in Sec.~\ref{limiting}. The importance of these results in the numerical estimation of entropies and densities of states is discussed in Sec.~\ref{practical}, followed by a summary and outlook in Sec.~\ref{conclusions}.

\section{Derivation} \label{derivation}

To begin the derivation, assume a purely classical framework is appropriate, and let $\vecG \equiv (\vecx,\vecp)$ be a point in the system phase space, where $\vecx,\vecp$ represent the canonically-conjugate position and momenta coordinates of all the particles, respectively. The system of interest has a Hamiltonian $H = H_\lambda (\vecx,\vecp)$ that depends parametrically on a predefined time-dependent function $\lambda(t)$. This function is an external parameter that ``switches'' from $\lambda(0) = A$ to $\lambda(\tau) = B$ in a given switching time $\tau$. In order to represent an isoenergetic process, one needs to model a suitable ``thermostat'' that exchanges heat with the system so as to precisely counterbalance the work done by the external parameter $\lambda$. One possible way of achieving this is to modify Hamilton's equations of motion by introducing an artificial ``force'' field $\vecF \equiv (\vecF_x,\vecF_p)$ on $\vecG$ as
\begin{align}
  \dot{\vecx} & = \frac{\partial H}{\partial \vecp} + \vecF_x(\vecG), \label{motion1}\\
  \dot{\vecp} & = -\frac{\partial H}{\partial \vecx} + \vecF_p(\vecG), \label{motion2}
\end{align}
and demand the constancy of the system energy through
\begin{equation} \label{dhdt}
  \frac{dH}{dt} = \nabla H \cdot \dot{\vecG} + \frac{\partial H}{\partial \lambda} \dot{\lambda} = 0,
\end{equation}
where $\nabla \equiv \partial / \partial \vecG$. Using the modified equations of motion, one immediately sees that the condition expressed by Eq.~(\ref{dhdt}) is fulfilled by any vector $\vecF$ satisfying $\nabla H \cdot \vecF = - \dot{\lambda} \, \partial H / \partial \lambda$, i.e. the desired force field is given in general by
\begin{equation} \label{F-general}
  \vecF = - \dot{\lambda} \, \frac{\vecX}{\vecX\cdot \nabla H} \frac{\partial H}{\partial \lambda},
\end{equation}
where $\vecX = \vecX(\vecG)$ is an arbitrary vector field not completely perpendicular to $\nabla H$. Particular examples of $\vecX$ satisfying this condition are $\vecX = (0,\partial H/\partial \vecp)$, and $\vecX = \nabla H$ itself. The former resembles the ``Gaussian'' thermostats widely used in the literature to model and simulate nonequilibrium processes (see e.g. Refs. \cite{evans-book,morriss98} for reviews), and will soon be discussed in more detail. In general, however, Eq.~(\ref{F-general}) provides a more flexible recipe particularly suited for parameter-dependent Hamiltonians.

The time-dependent ensemble density for the above non-Hamiltonian dynamics will now be derived. By direct integration of Liouville's equation in Lagrangian form (cf. Eq.~(3.3.8) in Ref.~\cite{evans-book}), one obtains the general solution
\begin{equation}
  \rho_t(\vecG_t) = \rho_0(\vecG_0) \, e^{-t\,\overline{\Lambda}_t(\vecG_t)} \label{L-def1},
\end{equation}
where
\begin{equation}
  \overline{\Lambda}_t(\vecG_t) \equiv \frac{1}{t} \int_0^t \! \! ds \, \Lambda(\vecG_s) \label{L-def2}
\end{equation}
is the time average of the ``phase space compression factor'' $\Lambda(\vecG) \equiv \nabla \cdot \dot{\vecG}$ along the trajectory that connects $\vecG_0$ to $\vecG_t$.
The notation $\overline{\Lambda}_t(\vecG_{t})$ implies that we are looking at the above functional of the phase space {\em trajectory} $\{ \vecG_s \}$, $s:[0,t]$ as a function of the {\em endpoint} $\vecG_t$, this being possible due to the deterministic character of these trajectories.
For a system evolving under the isoenergetic equations of motion Eqs.~(\ref{motion1})-(\ref{motion2}), with $\vecF$ given by Eq.~(\ref{F-general}), one has
\begin{equation} \label{L-specific}
  \overline{\Lambda}_t(\vecG_t) = - \frac{1}{t} \int_0^t \! \! ds \, \dot{\lambda} \, \nabla \cdot \left( \frac{\vecX}{\vecX\cdot \nabla H} \frac{\partial H}{\partial \lambda} \right)
\end{equation}
so the desired density at $\tau$ is completely determined by Eqs.~(\ref{L-def1}), (\ref{L-specific}) and the initial condition
\begin{equation} \label{initcond}
  \rho_0(\vecG_0) = \frac{\delta [E - H_A(\vecG_0)]}{\Omega_A(E)},
\end{equation}
which is just the microcanonical distribution. Here $\delta$ is the Dirac delta function, and $\Omega_\lambda(E) \equiv \int d\vecG \, \delta [E - H_\lambda(\vecG)]$ is the {\em density of states} at the external parameter $\lambda$.

Consider now the following average over all realizations of a switching process that takes $\lambda(t)$ from $A$ to $B$ in $\tau$ units of time:
\begin{align}
  \lang e^{\tau \overline{\Lambda}_{\tau}} \rang & = \int d\vecG_{\tau} \, \rho_{\tau}(\vecG_{\tau}) \, e^{\tau \overline{\Lambda}_{\tau}(\vecG_{\tau})} \nonumber \\
                         & = \int d\vecG_{\tau} \, \frac{\delta [E - H_A(\vecG_0)]}{\Omega_A(E)}. \label{el1}
\end{align}
In the second line above, Eqs.~(\ref{L-def1}) and (\ref{initcond}) have been used. Since the system Hamiltonian is constant along any trajectory generated by Eqs.~(\ref{motion1})-(\ref{motion2}) [cf. Eq.~(\ref{dhdt})], in particular $H_A(\vecG_0) = H_B(\vecG_{\tau})$, it follows from Eq.~(\ref{el1}) that
\begin{equation} \label{identity-exact}
  e^{\Delta S(E)} = \lang e^{\tau \overline{\Lambda}_{\tau}} \rang,
\end{equation}
where $\overline{\Lambda}_{\tau}$ is given by Eq.~(\ref{L-specific}), and $\Delta S(E) \equiv \ln [ \Omega_B(E)/\Omega_A(E) ]$ is the entropy difference between the thermodynamic states $A$ and $B$. The identity above, along with its quasistatic version [Eq.~(\ref{dS-int})], are the central results of the present paper [see also Eq.~(\ref{identity-particular})]. Some computational aspects of these results will be discussed shortly.

\section{Limiting cases} \label{limiting}

Let us now study the quasistatic limit of Eq.~(\ref{identity-exact}), i.e. the limiting case where $\lambda(t)$ changes infinitely slowly from $A$ to $B$. In this case, by a slight extension of the ``adiabatic ergodic hypothesis'' \cite{rugh01}, one expects that the dynamical state of the system ($\vecG_s$) spends a sufficiently long time sweeping a nearly constant-energy, constant-$\lambda$ surface $E=H_\lambda (\vecG_s)$ so that one can invoke the approximation
\begin{equation}
  \frac{1}{\tau} \int_0^{\tau} \! \! ds \, \Lambda(\vecG_s) = \frac{1}{\tau} \int_0^{\tau} \! \! ds \, \lang \Lambda \rang_{E,\lambda(s)}, \quad \text{(quasistatic)} \nonumber
\end{equation}
where the subscripted angle brackets denote a {\em microcanonical} ensemble average at constant $E$ and $\lambda(s)$. Accordingly, the time average $\overline{\Lambda}_{\tau}$ is independent of the specific realization of the switching process, and it follows after a simple change of integration variables that Eq.~(\ref{identity-exact}) reduces to the following new thermodynamic integration formula:
\begin{equation} \label{dS-int}
  \Delta S(E) = - \int_A^B \! \! d\lambda \, \lang \nabla \cdot \left( \frac{\vecX}{\vecX\cdot\nabla H} \frac{\partial H}{\partial \lambda} \right) \rang_{E,\lambda}.
\end{equation}
Recall that the Jarzynski equality also reduces to a thermodynamic integration formula in the limit of quasistatic processes, viz. $\Delta F(T) = \int_A^B \! d\lambda \, \lang \partial H / \partial \lambda \rang_\lambda$ \cite{jarzynski97-prl}, a result that can be easily derived by standard statistical mechanics. The existence of an arbitrary (see above) vector field in Eq.~(\ref{dS-int}) is analogous to a generalization of the thermodynamic integration formula for $\Delta F$ also derived by Jarzynski \cite{jarzynski02}. 

An independent proof of Eq.~(\ref{dS-int}) based on ergodic manipulations \cite{khinchin49} is now provided as a consistency check. Starting from $\Delta S(E) = \int_A^B \! d\lambda \, (\partial S/\partial \lambda)$ and $S_\lambda(E) \equiv \ln \Omega_\lambda (E)$, one has
\begin{align}
  \frac{\partial}{\partial \lambda} \ln \Omega & = \frac{1}{\Omega} \frac{\partial}{\partial \lambda} \int \! d\vecG \, \delta[E-H_\lambda (\vecG)] \nonumber \\
    & = - \frac{1}{\Omega} \frac{\partial}{\partial E} \int \! d\vecG \, \delta[E-H_\lambda (\vecG)] \, \frac{\partial H}{\partial \lambda} \nonumber \\
    & = - \frac{1}{\Omega} \frac{\partial}{\partial E} \oint_\Sigma \! \frac{da}{|\nabla H|} \frac{\partial H}{\partial \lambda}, \label{partialS}
\end{align}
where $\Sigma$ is the surface $E=H_\lambda (\vecG)$, and $da$ is an infinitesimal area element of this surface. Now let $\vecX=\vecX(\vecG)$ be a vector field not completely perpendicular to $d\veca \equiv da \, \nabla H / |\nabla H|$. Then 
\begin{equation}
\frac{da}{|\nabla H|} = d\veca \cdot \frac{\vecX}{\vecX \cdot \nabla H}, \nonumber
\end{equation}
so that, by the divergence theorem and a subsequent energy differentiation, Eq.~(\ref{partialS}) becomes
\begin{equation}
  \frac{\partial}{\partial \lambda} \ln \Omega = - \lang \nabla \cdot \left( \frac{\vecX}{\vecX\cdot\nabla H} \frac{\partial H}{\partial \lambda} \right) \rang_{E,\lambda}, \nonumber
\end{equation}
which coincides with Eq.~(\ref{dS-int}) upon integration, q.e.d. 

Although Eqs.~(\ref{identity-exact}) and (\ref{dS-int}) are exact and can in principle be directly utilized in numerical simulations (Sec.~\ref{practical}), it is interesting to investigate their limit in the case of large systems. Consider first the integrand of Eq.~(\ref{L-specific}). One has, exactly,
\begin{equation} \label{T-introduced}
  \nabla \cdot \left( \frac{\vecX}{\vecX\cdot \nabla H} \frac{\partial H}{\partial \lambda} \right) = \frac{1}{\mathcal{T}} \frac{\partial H}{\partial \lambda} 
  + \frac{\vecX\cdot \nabla(\partial H /\partial \lambda)}{\vecX\cdot \nabla H},
\end{equation}
where
\begin{equation}
  \frac{1}{\mathcal{T}} \equiv \nabla \cdot \frac{\vecX}{\vecX\cdot \nabla H} \nonumber
\end{equation}
is an inverse ``instantaneous temperature'' that coincides with the inverse of a recently derived expression for the microcanonical temperature \cite{rugh97,rugh98} upon ensemble-averaging. Assume now that, as expected, $\mathcal{T}$ is intensive. Two possibilities of interest arise: (a) $\lambda$ couples to a selected number of particles and $\vecX$ thermostats all the particles, or (b) $\lambda$ couples to all the particles and $\vecX$ thermostats any number of particles\footnote{The relevant case (a) was suggested by Gavin Crooks (personal communication).}. Under these conditions and as the total number of particles is increased, the second term in the r.h.s. of Eq.~(\ref{T-introduced}) either (a) vanishes or (b) becomes negligible in comparison to the first one. For large enough systems, therefore, Eq.~(\ref{L-specific}) can be approximated as
\begin{equation}
  \overline{\Lambda}_{\tau} = - \frac{1}{\tau } \int_0^{\tau} \! \! dt \, \dot{\lambda} \frac{\partial H}{\partial \lambda} \frac{1}{\mathcal{T}} , \quad \text{(large systems)} \nonumber
\end{equation}
and the nonequilibrium identity Eq.~(\ref{identity-exact}) can be rewritten as
\begin{equation}
  \lang e^{ - \int_A^B \! dW/ \mathcal{T}} \rang = e^{\Delta S(E)}, \quad \text{(large systems)} \nonumber
\end{equation}
where $\dot{\lambda} \, \partial H / \partial \lambda$ has been identified with the rate of work input due to the external parameter $\lambda$ \cite{jarzynski02-lecnotes} [see also Eq.~(\ref{dhdt})], i.e.  
\begin{equation}
  dW \equiv \frac{\partial H} { \partial \lambda} \, d\lambda .
\end{equation}
By invoking Jensen's inequality $\lang e^x \rang \geq e^{\lang x \rang}$ \cite{chandler87-ineq}, the above result can also be recast in a form that resembles the second law of thermodynamics for isoenergetic processes (i.e. the Clausius inequality with $dQ=-dW$), namely 
\begin{equation} \label{dS-thermo}
  \Delta S(E) \geq - \lang  \int_A^B \frac{dW}{\mathcal{T}} \rang, \quad \text{(large systems)}
\end{equation}
where, as in the case of the JE, the angle brackets can presumably be dropped in the thermodynamic limit\footnote{An additional microscopic connection (not investigated here) between $\mathcal{T}$ and the instantaneous reservoir temperature $T$ is necessary before the inequality (\ref{dS-thermo}) reduces to a statement of the second law of thermodynamics.}. It is not difficult to verify, however, that in the case of {\em quasistatic} processes for sufficiently large systems, one obtains from either Eq.~(\ref{dS-int}) or Eq.~(\ref{dS-thermo}) the thermodynamic statement
\begin{equation}
  \Delta S(E) = -\int_A^B \frac{dW}{T}, \quad \text{(large systems, quasistatic)} \nonumber
\end{equation}
where $1/T \equiv \lang 1/\mathcal{T} \rang_{E,\lambda}$ is the inverse equilibrium temperature of the system, and hence of the reservoir along the isoenergetic path.

\section{Utility and practical aspects} \label{practical}

It is natural to inquire about the utility of the above results, since they were derived after the admittedly artificial equations of motion described by Eqs.~(\ref{motion1})-(\ref{motion2}). Though in physical experiments it is very unlikely that one will ever encounter such an isoenergetic setting, this is not the case in numerical simulations (see e.g. \cite{evans-book,morriss98}). These results should therefore be of greatest interest for the computational community, but even in this case one might ask why the computation of $S(E)$ or $\Omega(E)$ is at all relevant, since quantities such as free energies at constant temperature/pressure have a more direct connection with experiments.

Although an immediate answer to the above question can be found in the context of phase transitions of finite or ``small'' systems \cite{gross00,gross02}, I will briefly describe how one can recover important thermodynamic quantities in the {\em isothermal} ensemble given the knowledge of $S(E)$ or $\Omega(E) = e^{S(E)}$ over a range of energy values. How to efficiently obtain entropy and density of states at more than a single energy value will be discussed shortly.

First, notice that the canonical ensemble average at temperature $T$ of any energy-dependent observable $f(E)$ can be written as
\begin{equation} \label{energy-average}
  \lang f(E) \rang_T = \frac{\int_0^{\infty} \! dE' \, \Omega(E') \, e^{-E'/T} f(E')} {\int_0^{\infty} \! dE' \, \Omega(E') \, e^{-E'/T}}.
\end{equation}
Therefore, knowledge of the density of states over the relevant range of energies where $\Omega(E') \, e^{-E'/T}$ is concentrated allows one to calculate observables such as average potential energy, heat capacities, and free energy differences by simple quadrature. In fact, $\Omega(E)$ plays a central role in the so-called ``flat histogram'' Monte Carlo methods \cite{wang01a,wang01b}, and is used precisely as described above to estimate such observables. It is worth mentioning that the present nonequilibrium method shares an additional feature with flat histogram methods, which have been introduced with the goal of overcoming energy barriers in finite-temperature simulations. An estimate based on Eq.~(\ref{identity-exact}) also has the potential of sampling the phase space without becoming trapped in energy basins defined by barriers much greater than $T$ (recall that $k_B=1$ in the present units). This follows from the ability to construct a switching process that changes the strength of the interaction among the particles, as described in the next paragraph, so that during the course of the simulation the dynamics enjoys much lower energy barriers and hence greater mobility.

The results derived in this work can be made highly efficient by borrowing a scaling idea originally developed for the adiabatic switching method \cite{dekoning99}. In the present language, the scaling idea allows one to obtain entropy differences or ratios of densities of states between the system of interest and a reference system over a wide range of energies from the data obtained at a {\em single} energy value $E$. In fact, assuming that the reference system is an ideal gas, i.e. the parameter-dependent Hamiltonian is $H_\lambda(\vecx,\vecp)=\vecp^2/2+\lambda \, U(\vecx)$ with $\lambda=\lambda(t)$ switching from $0$ to $1$ in $\tau$ units of time, by rearranging the expression $\Omega_\lambda(E) = \int \! d\vecx d\vecp \, \delta[E-\vecp^2/2-\lambda\,U(\vecx)]$ one can easily derive the following scaling relation:
\begin{equation} \label{scaling}
  S_1(E/\lambda) - S_0(E) = \Delta S_\lambda(E) - \frac{Nd-2}{2} \, \ln \lambda.
\end{equation}
Here $S_1(E/\lambda)$ is the entropy of the system of interest at the energy value $E/\lambda$, $S_0(E)$ is the (known) ideal gas entropy, and the difference $\Delta S_\lambda(E)\equiv S_\lambda (E) - S_0(E)$ is obtained either directly from Eq.~(\ref{dS-int}) with the upper integration limit replaced by $\lambda$, or by recording the intermediate values of the average in Eq.~(\ref{identity-exact}) at the instant $t$ corresponding to the desired $\lambda=\lambda(t)$. Note finally that the expressions derived herein for $\Delta S(E)$ can equally well be used for estimating ratios of density of states, $\Omega_B(E)/\Omega_A(E)$, without resorting to the traditional histograms.

It follows from the above scaling identity that, provided the entropy/density of states of the reference system is known with great accuracy, the numeric value of $S(E)$ or $\Omega(E)$ can be obtained for a wide range of energies from a single isoenergetic simulation and, in particular, one is able to use Eq.~(\ref{energy-average}) to estimate various thermodynamic quantities of interest over a range of temperature values. In practice, of course, one has to choose the range of $E$ judiciously so that the estimates of $\Omega(E)$ are reasonably accurate in the range where the integrand of Eq.~(\ref{energy-average}) is greatest.

Several other properties enjoyed by the present results will now be discussed. In contrast to the ``adiabatic switching'' method for computing entropy differences \cite{watanabe90}, the exact result in Eq.~(\ref{identity-exact}) does not suffer from the systematic errors associated with a finite switching time \cite{frenkel02}. Moreover, since neither Eq.~(\ref{identity-exact}) nor (\ref{dS-int}) relies on a single dynamical trajectory, these equations are trivially parallelizable. Isoenergetic equations of motion similar to those adopted in this work have been extensively used in the literature as means of simulating nonequilibrium processes \cite{evans-book,evans02}, and the general recipe provided in Eqs.~(\ref{motion1})-(\ref{motion2}) and (\ref{F-general}) should not pose any additional technical difficulty. Nonetheless, a considerable operational simplification is achieved when $\vecX = (0,\partial H / \partial \vecp) = (0,\vecp)$, where it is assumed that $H_\lambda(\vecx,\vecp)=\vecp^2/2 + U_\lambda(\vecx)$. In this case, the thermostatting mechanism reduces to a simple velocity-dependent force $\vecF_p = -\dot{\lambda} (\partial U / \partial \lambda) \vecp/ \vecp^2$, and Eq.~(\ref{identity-exact}) yields {\em exactly}
\begin{equation} \label{identity-particular}
  e^{\Delta S(E)} = \lang e^{ - \int_A^B \! dW/ \mathcal{T}} \rang,
\end{equation}
where $dW=dt\, \dot{\lambda}( \partial U /\partial \lambda)$ is the infinitesimal amount of work performed on the system, and $1/\mathcal{T} = (Nd-2)/\vecp^2$ is the inverse instantaneous temperature defined in Sec.~\ref{limiting} ($d$ is the number of spatial dimensions and $N$ is the number of particles). The above equation has the advantage of involving less abstract and more physically sound quantities, while remaining as rigorous as the more general case of Eq.~(\ref{identity-exact}).

\section{Conclusions} \label{conclusions}

In summary, the present contribution has introduced two fundamental nonequilibrium and equilibrium equalities for entropy differences (or ratios of densities of states), Eqs. (\ref{identity-exact}) and (\ref{dS-int}) respectively [see also Eq.~(\ref{identity-particular})]. The former is a nontrivial extension of the Jarzynski equality [Eq.~(\ref{jarzynski})] to the microcanonical ensemble, which was made possible through the construction of a suitable set of non-Hamiltonian equations of motion. This particular extension circumvents the mathematical difficulties that one encounters by adapting straightforward strategies that have been successful in other ensembles \cite{crooks98,park04}. The latter is the entropic analog of the well-known thermodynamic integration formula for free energies.  

The utility of these results was discussed in Sec.~\ref{practical}, where it was remarked that the artificial character of the isoenergetic equations of motion described by Eqs.~(\ref{motion1})-(\ref{motion2}) causes such results to be of greatest interest in numerical simulations. In this context, the identities derived in this work allow one to compute $S(E)$ (entropy at a given energy) or $\Omega(E)$ (density of states) for a wide range of energies from a {\em single} isoenergetic simulation, as discussed in connection with Eq.~(\ref{scaling}). Knowledge of these quantities per se is of great interest in the study of phase transitions of finite or ``small'' systems \cite{gross00,gross02}, but can also be used to recover important observables in other ensembles (e.g. isothermal), as evidenced by Eq.~(\ref{energy-average}).

Since Eqs.~(\ref{identity-exact}) and (\ref{dS-int}) share in common an arbitrariness through the vector field $\vecX$, a question that deserves further investigation is whether one can find an optimal form for $\vecX$ that maximizes computational efficiency.

\begin{acknowledgments}

The author is pleased to thank Chris Jarzynski for the illuminating discussions and suggestions. Discussions and/or correspondences with Gavin Crooks, Jimmie Doll, Heather Partner and Tom Woolf are also acknowledged. This research was supported by the Department of Energy, under contract W-7405-ENG-36, as well as under grant DE-FG02-03ER46074.

\end{acknowledgments}

\end{document}